\documentclass[%
 reprint,
 amsmath,amssymb,
 aps,
]{revtex4-2}

\usepackage{xcolor}

\usepackage{graphicx}
\usepackage{dcolumn}
\usepackage{bm}

\usepackage[normalem]{ulem}

\begin{document}

\newcommand{\U}{\mathrm{^{235}U}} 
\newcommand{\Pu}{\mathrm{^{239}Pu}}

\preprint{APS/123-QED}

\title{Reactor Antineutrino Spectral ``Bump": Cumulative Fission Yields of Irradiated $^{235}$U and $^{239}$Pu Measured by HPGe Gamma-Ray Spectroscopy }

\author{Samuel Kim}
\affiliation{%
 Brookhaven National Laboratory, Temple University
}%
\author{C. J. Martoff}%
 \affiliation{%
 Temple University
}%
\author{Michael Dion}
\altaffiliation[Currently at ]{Thomas Jefferson National Accelerator Facility}
\affiliation{%
 Oak Ridge National Laboratory
}%

\author{David Glasgow}%
\affiliation{%
 Oak Ridge National Laboratory\\
}%


\date{\today}

\begin{abstract}
Recent measurements of the reactor antineutrino emission show that there exists a spectral excess (the ``bump") in the 5-7 MeV region when compared to the Huber-Muller prediction based on the conversion method. Analysis within an alternate prediction technique, the summation method, suggests that the bump could be due to excess contributions from a certain few of the $\beta$-decaying fission products. However, it has been shown that when updated fission yield values are used in the summation method, the predicted excess vanishes. In the present preliminary study, fission yields for nuclides suspected of causing the neutrino spectral bump are investigated using gamma-ray spectroscopy of $\mathrm{^{235}U}$ and $\mathrm{^{239}Pu}$ samples freshly irradiated using the High Flux Isotope Reactor.  For several of the suspect nuclides, the derived fission yields are consistent with JEFF3.3 fission yield library. The exception is the case of $\mathrm{^{140}Cs}$ from $\mathrm{^{239}Pu}$, where the discrepancy between the fitted and expected values suggests a potential error in the fission yield library. This highlights the importance of using accurate nuclear data libraries in the analysis of the reactor antineutrino spectra, and the need for ongoing efforts to improve these libraries. 
\end{abstract}

\maketitle


\section{\label{sec:level1}Introduction\protect\\} 


Nuclear reactors are intense sources of electron antineutrinos, and are therefore widely used to study the complex properties of these intriguing particles. From one fission, approximately 6 $\bar\nu_e$ are produced, and a 1 GW thermal reactor emits about 10$^{20}$ $\bar\nu_e$ per second \cite{hayes2016reactor}. Fission fragments are neutron-rich, resulting in $\beta$ decays. Recent large-scale anti-neutrino spectral measurements \cite{KIM201593, abe2012indication} show that there is a spectral bump in the 5 to 7 MeV region of $\bar\nu_e$ that is not predicted by the $\beta$-conversion method of predicting the expected neutrino spectrum from measured beta spectra (Huber-Muller method).  The aggregated beta spectrum is made up of thousands of decay channels with different end point beta energies. Conversion to an antineutrino spectrum is performed by fitting the measured electron spectrum with a superposition of 30 end-point beta energies and using the kinematics of $\beta$-decay to obtain the corresponding neutrino spectra \cite{Muller-PRC054615}. 

In the summation method, an alternate approach is used to evaluate the $\bar\nu_e$ spectrum.   Nuclear data files such as Evaluated Nuclear Data File (ENDF) library and Joint Evaluated Fission and Fusion File (JEFF) library are used to estimate the associated neutrino spectrum using all the relevant tabulated fission yields and $\beta$-decay parameters. 
Based on the ENDF/B-VII library, the summation method suggests that the spectral bump could be due to yields in excess of the eight particular $\beta$-decaying fission products, which give a combined 42\% of the total decay rate in the $\beta$-energy region of 4 to 6 MeV ($\bar\nu_e$ energy region of 5 to 7 MeV) \cite{dwyer2015spectral}. Table \ref{tab:table1} lists the same eight fission products discussed in Ref.\cite{dwyer2015spectral} as the primary contributors to the spectral bump.   
\begin{table}[htpb]
\caption{\label{tab:table1}%
Decay data for 8 nuclides singled out in Ref. \cite{dwyer2015spectral} from the ENDF/B-VIII decay data sublibrary, including the decay chain gamma-ray with the strongest intensity selected for the present analysis.  Uncertainty is given in the parenthesis.  
}
\begin{ruledtabular}
\begin{tabular}{cccc}
\textrm{Isotope}&
\textrm{Half life (s)}&
\textrm{Gamma Energy (keV)}&
\textrm{Intensity}\\
\colrule
$\mathrm{^{93}Rb}$ & 5.84(2) & 432.61(3) & 0.202(14)\\
$\mathrm{^{100}Nb}$& 1.4(2) & 535.666(14) & 0.46(6)\\
$\mathrm{^{140}Cs}$ & 63.7(3) & 602.25(5) & 0.53(3)\\
$\mathrm{^{95}Sr}$ & 23.90(14) & 685.6 & 0.226\\
$\mathrm{^{92}Rb}$ & 4.49(3) & 814.98(3) & 0.032(4)\\
$\mathrm{^{96}Y}$ & 5.34(5) & 1750.4(2) & 0.0235(24)\\
$\mathrm{^{97}Y}$ & 3.75(3) & 3287.6(4) & 0.181(19)\\
$\mathrm{^{142}Cs}$ & 1.68(14) & 359.598(14) & 0.27(3)\\
\end{tabular}
\end{ruledtabular}
\end{table}

In a follow-up study, Sonzogni et al.\cite{sonzogni2016effects} demonstrate that reproduction of the bump based on the summation method is due to errors in the fission yield values contained in ENDF/B-VII library.  When corrected and improved fission yield values are used, no excess contributions from the eight nuclides are observed. 
In the present work, fission is induced in \(\U \)   and  \(\Pu \)    samples by neutron irradiation in the High Flux Isotope Reactor (HFIR), and the resulting gamma-ray spectra are measured by a high purity germanium (HPGe) detector after rapid transport out of the core. The measured spectra are compared to predictions based on data from JEFF3.3 fission yield library and the ENDF/B-VIII decay data sublibrary. 

\section{Experiment}

The \(\U \) sample consists of 252.72 nanograms of natural uranium nitrate in an Inductively Coupled Plasma calibration solution.  The \(\Pu \) sample consists of 301.3 nanogram of National Institute of Standards and Technology (NIST) Certified Reference Material (CRM-137).  The samples are irradiated using the PT-2 pneumatic tube of the HFIR at the Neutron Activation Analysis laboratory (NAA) of Oak Ridge national Laboratory.  The measured thermal and epithermal neutron fluxes at the irradiation location are 4.59$\times$10$^{13}$ n/cm$^{2}$/sec and 1.96E$\times$10$^{11}$ n/cm$^{2}$/sec respectively for \(\U \), and 4.43$\times$10$^{13}$ n/cm$^{2}$/sec and 3.24$\times$10$^{11}$ n/cm$^{2}$/sec respectively for \(\Pu \).  The neutron fluxes are measured using manganese and gold activation foils. 

Each sample is irradiated for 30 seconds, and then transported to the detector chamber using the pneumatic tube transfer system \cite{reactor2015user} which introduces a 20-second delay prior to the gamma-ray measurement.  This delay is problematic for the short-lived $^{97}$Y and $^{142}$Cs.  Future work is planned to reduce the delay and improve detection sensitivity.  Fig. 1 shows the measured gamma-ray spectra of the irradiated \(\U \)   and  \(\Pu \).
\begin{figure*}[hbt!]
\includegraphics[width=\linewidth]{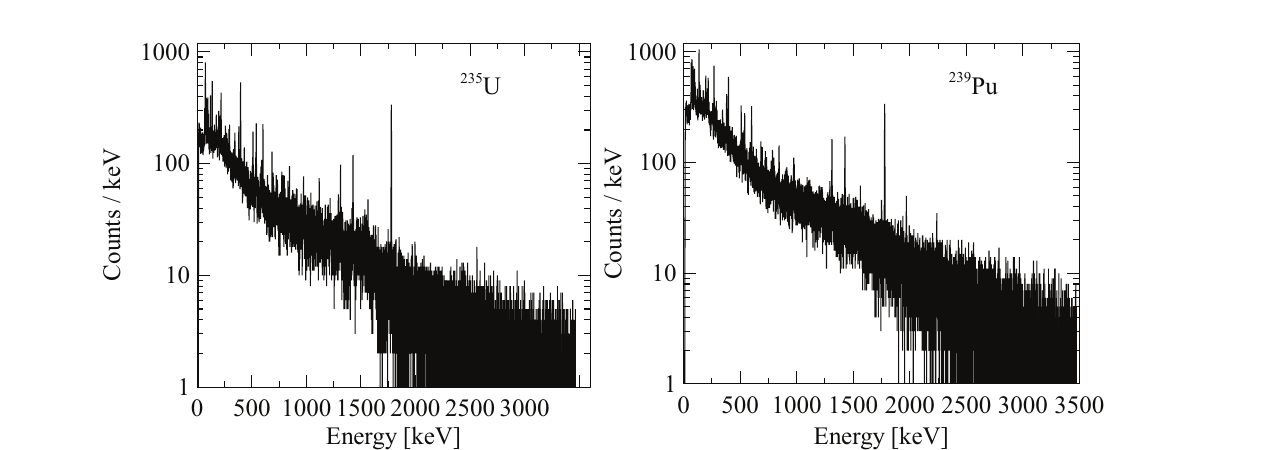}
\caption{\label{fig:fig1} Measured gamma-ray spectra from freshly irradiated \(\U \)   and  \(\Pu \) are plotted. See text for details.}
\end{figure*}

The gamma rays are measured with a 44\% relative efficiency,  ORTEC p-type coaxial HPGe detector with an aluminum end cap.   Each sample is placed at 33 cm above the detector and measured for 30 seconds. For the analysis presented in this work, only the $\beta$-decay path for the parent-daughter chains is used, neglecting $\beta$-delayed neutron emission channels.  

\section{Calculated gamma rays}
The expected gamma-ray yield calculation starts by determining the number of \(\U \) and \(\Pu \) nuclides initially present in the sample from the sample mass (m), Avogadro’s number ($N_A$) and the molar weight (M). Total fission production and decay of the gamma emitter during the irradiation ($N_\textrm{fd}$) is determined using Eq. (\ref{eq1}).   

\begin{equation}
 N_\textrm{fd} = 
 \textrm{IFY}\, \sigma_f\, \phi\,   \frac{mN_A}{M} (1-e^{-\lambda t})
\label{eq1}
\end{equation}
The equation includes the independent fission yield (IFY) of a specific nuclide, the thermal neutron cross section ($\sigma_f$) and the thermal neutron flux ($\phi$) and the irradiation \cite{knowles2016generalized}.  The IFY is obtained from the JEFF3.3 library, and the neutron cross section is based on the ENDF/B-VIII.0 neutron cross section standard sublibrary. The JEFF3.3 and ENDF/B-VIII.0 fission yield libraries contain different IFY values for certain nuclides. This is demonstrated using the $\mathrm{^{140}Cs}$ decay chain in Table \ref{tab:table2}. In this example, JEFF3.3 does not have IFY for $\mathrm{^{140}Sb}$, so the IFY value from ENDF/B-VIII.0 is used instead in our analysis. Thus, $N_\textrm{fd}$ will be different depending on the fission library used.  

\begin{table*}[htp]
\renewcommand*{\arraystretch}{1.3}
\caption{\label{tab:table2}%
Examples from the $\mathrm{^{140}Cs}$ decay chain, showing the differing IFY of \(\U \) and \(\Pu \)  from JEFF3.3 and ENDF/B-VIII.0  fission yield libraries.  Uncertainty of each IFY is indicated in the parenthesis.
}
\begin{ruledtabular}
\begin{tabular}{cccccc}
\textrm{IFY (\(\U \))}&
\textrm{$\mathrm{^{140}Sb}$}&
\textrm{$\mathrm{^{140}Te}$}&
\textrm{$\mathrm{^{140}I}$}&
\textrm{$\mathrm{^{140}Xe}$}&
\textrm{$\mathrm{^{140}Cs}$}\\
\colrule
\textrm{JEFF3.3} & \textrm{No data} & 6.57E-08  (2.26E-08) & 3.03E-04  (1.03E-04) & 1.25E-02  (3.10E-03) & 1.84E-02  (3.85E-03)\\
\textrm{ENDF/B-VIII.0}& 2.82E-09  (1.81E-09) & 9.04E-06  (5.78E-06) & 1.11E-03  (7.13E-04) & 2.59E-02  (1.04E-03) & 3.05E-02  (1.83E-03) \\
\hline \hline 
\textrm{IFY (\(\Pu \))}&
\textrm{$\mathrm{^{140}Sb}$}&
\textrm{$\mathrm{^{140}Te}$}&
\textrm{$\mathrm{^{140}I}$}&
\textrm{$\mathrm{^{140}Xe}$}&
\textrm{$\mathrm{^{140}Cs}$}\\
\colrule
\textrm{JEFF3.3} & \textrm{No data} & 2.33E-07  (8.06E-08) & 4.77E-04  (1.63E-04) & 1.83E-02  (4.06E-03) & 2.18E-02  (4.52E-03)\\
\textrm{ENDF/B-VIII.0}& 5.61E-11  (3.59E-11) & 1.41E-06  (9.02E-07) & 5.94E-04  (3.80E-04) & 1.54E-02  (4.31E-04) & 2.28E-02  (3.64E-03) \\
\end{tabular}
\end{ruledtabular}
\end{table*}

In addition, $N_\textrm{fd}$ of each precursor of the gamma emitter needs to be determined and $\beta$-decayed to properly account for the total number of  the gamma emitter produced at the end of the irradiation.  Cumulative yield (CY) used in this study is described in Eq. (\ref{eq2}).

\begin{equation}
 CY_{i} =
 [N_\textrm{fd}]_{i} + \sum_{ij} Decay( [N_\textrm{fd}]_{j}  )
\label{eq2}
\end{equation}

The second term describes the total number of $i$ gamma emitter resulting from the $\beta$-decay of $j^{th}$ precursor of $i$ gamma emitter. Each decay chain leading from primary fission products to a gamma ray emitter measured in this study is described by a set of coupled linear differential equations describing the radioactive decays.  These equations are reformulated as a set of matrices and solved using Matlab.  The solution to each decay chain gives the number of gamma emitter resulting from the decay of its precursors during the 30-second irradiation. Resulting total CY is further decayed for 20 seconds modeling the RABBIT transportation delay.

Expected gamma-ray yields during the subsequent (delayed) 30-second measurement time are calculated as factors of the decayed CY, decay constant ($\lambda$), absolute efficiency ($\epsilon$) of the HPGe, and the gamma emission intensity ($I_{\gamma}$).
The yield calculation is aided by Radiation Intensity Calculator (RadICal), a python-based application developed by Pacific Northwest National Laboratory (PNNL) researchers \cite{robinson2015radicalc}. Its solutions are built on Laplace transform of the Bateman
equation.

The half lives of the nuclides investigated here are much shorter than the detector measurement time, therefore, it is necessary to decay-correct the measured peak counts for the count time. The ANSI standard for the correction factor is described in Ref.\cite{768889}. The largest uncertainty contribution comes from the uncertainties associated with IFY.  The relative uncertainty \cite{taylor1997introduction} of IFY from each nuclides is: $\mathrm{^{93}Rb}$(18\%), $\mathrm{^{100}Nb}$(35\%), $\mathrm{^{140}Cs}$(24\%), $\mathrm{^{95}Sr}$(10\%), $\mathrm{^{92}Rb}$(18\%), $\mathrm{^{96}Y}$(24\%), $\mathrm{^{97}Y}$(15\%) and $\mathrm{^{142}Cs}$(20\%).

\section{Measured gamma rays }

The energy and full width at half maximum (FWHM) calibrations of the HPGe detector have previously been determined by analyzing known gamma ray peaks.  
The absolute efficiency of the detector is estimated using the Geometry and Tracking\cite{GEANT4} simulation package.  In the simulation, 17 gamma ray energies are selected to cover the energy range from 50 keV to 3.5 MeV.  Each gamma-ray simulation was performed using 1E+6 photons to determine the efficiency of the detector at each photon energy. The detector model in GEANT4 includes all the details of detector construction, including a 0.1 cm thick aluminum window on the endcap of the detector and a 0.07 cm thick dead layer on the surface of the HPGe crystal. Dimensions of the HPGe were taken to be 6.5 cm in the diameter and 6.45 cm in the length based on published ORTEC documents\cite{Roth2018}.
According to the ANSI/IEEE standard 325 \cite{fairstein1996ieee}, the relative efficiency of an HPGe is defined by Eq. (\ref{eq3}).  The absolute efficiency of HPGe at 1.33 MeV is measured with a source to detector distance of 25 cm. The relative efficiency is the ratio of this HPGe absolute efficiency to the absolute efficiency of a 3-inch by 3-inch Na(Tl) at 1.33 MeV measured at 25 cm (1.2E-3). 

\begin{equation}
 \mathrm{Relative\, efficiency}  = 
 \frac{\mathrm{Absolute\, efficiency} }{1.2 \times 10^{-3}}
\label{eq3}
\end{equation}

\begin{figure}[htp!]
\includegraphics[width=\linewidth]{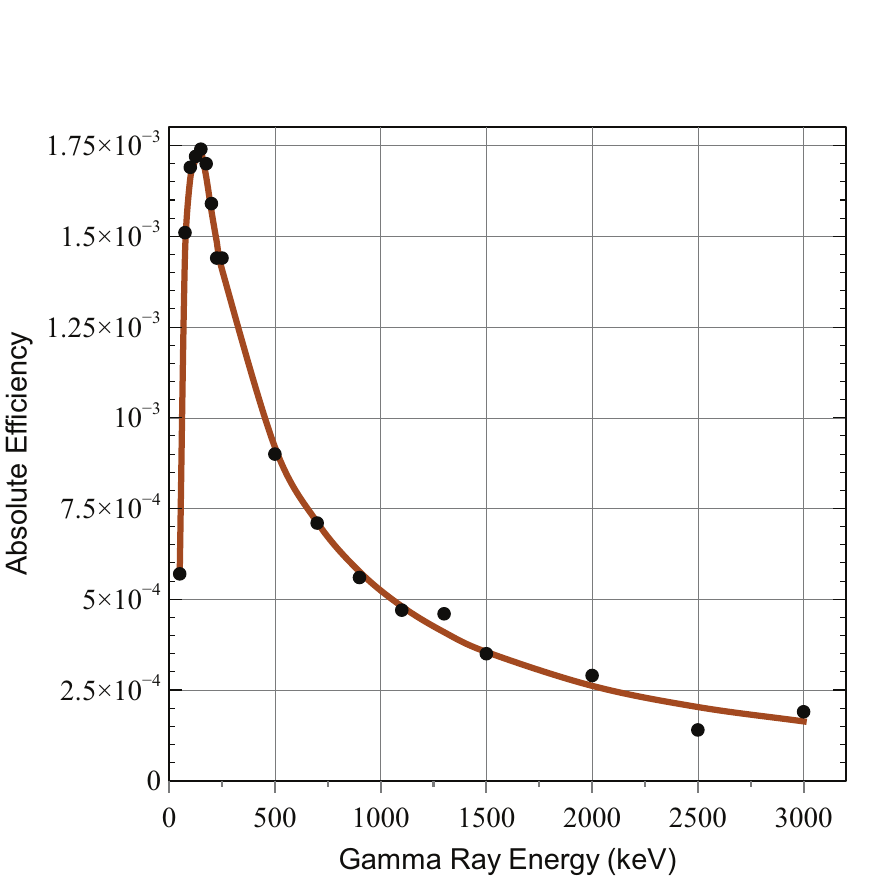}
\caption{\label{fig:fig2} GEANT4 simulated efficiency for the ORTEC P-type 44\% relative efficiency HPGe detector used here. The efficiency is fitted using the parametric equation given in the RADWARE program.
}
\end{figure}
To establish a benchmark, a GEANT4 simulation was performed for a point source placed at the standard distance of 25 cm from the detector. The absolute efficiency at 1333 keV was expected to be 5.3E-4 for a 44\% relative efficient HPGe \cite{ortec}. The simulated absolute efficiency was 5.9E-4(7.7E-5). Fig. \ref{fig:fig2} shows the simulated detector efficiency.  The efficiency is fitted using the parametric equation given in the RADWARE software package \cite{radfordradware}. Above 150 keV, efficiency is fitted with the parameters (D, E and F) in the form of:
\begin{equation}
 \mathrm{Efficiency}  = 
 e^{D+Ey+Fy^2}
\label{eq31}
\end{equation}
where $y = ln(E_{\gamma}/1000)$ and $E_{\gamma}$ is a gamma-ray energy in keV.

\begin{figure}[htp!]
\includegraphics[width=\linewidth]{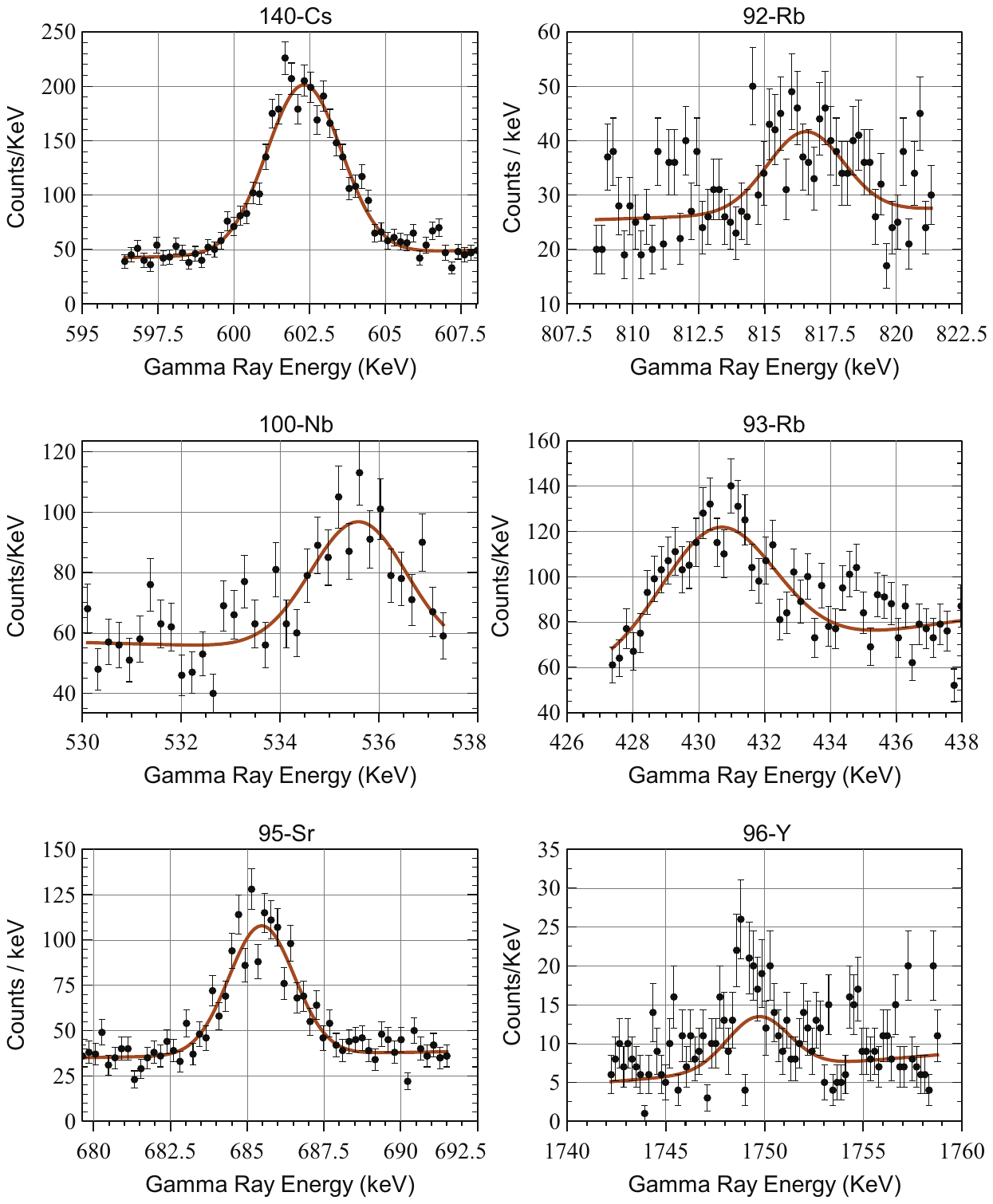}
\caption{\label{fig:fig3} Data and fits for the six gamma-ray peaks of interest from 235-U fission  are shown.  The measured spectra are shown by the circled  dots with 1-$\sigma$ uncertainties, and the fits by solid lines. Due to low yield, environmental backgrounds, and Compton scattering from higher energy photons, gamma ray peaks from $\mathrm{^{97}Y}$  and $\mathrm{^{142}Cs}$ are not detectable, and omitted in this figure. }
\end{figure}

\begin{figure}[htp]
\includegraphics[width=\linewidth]{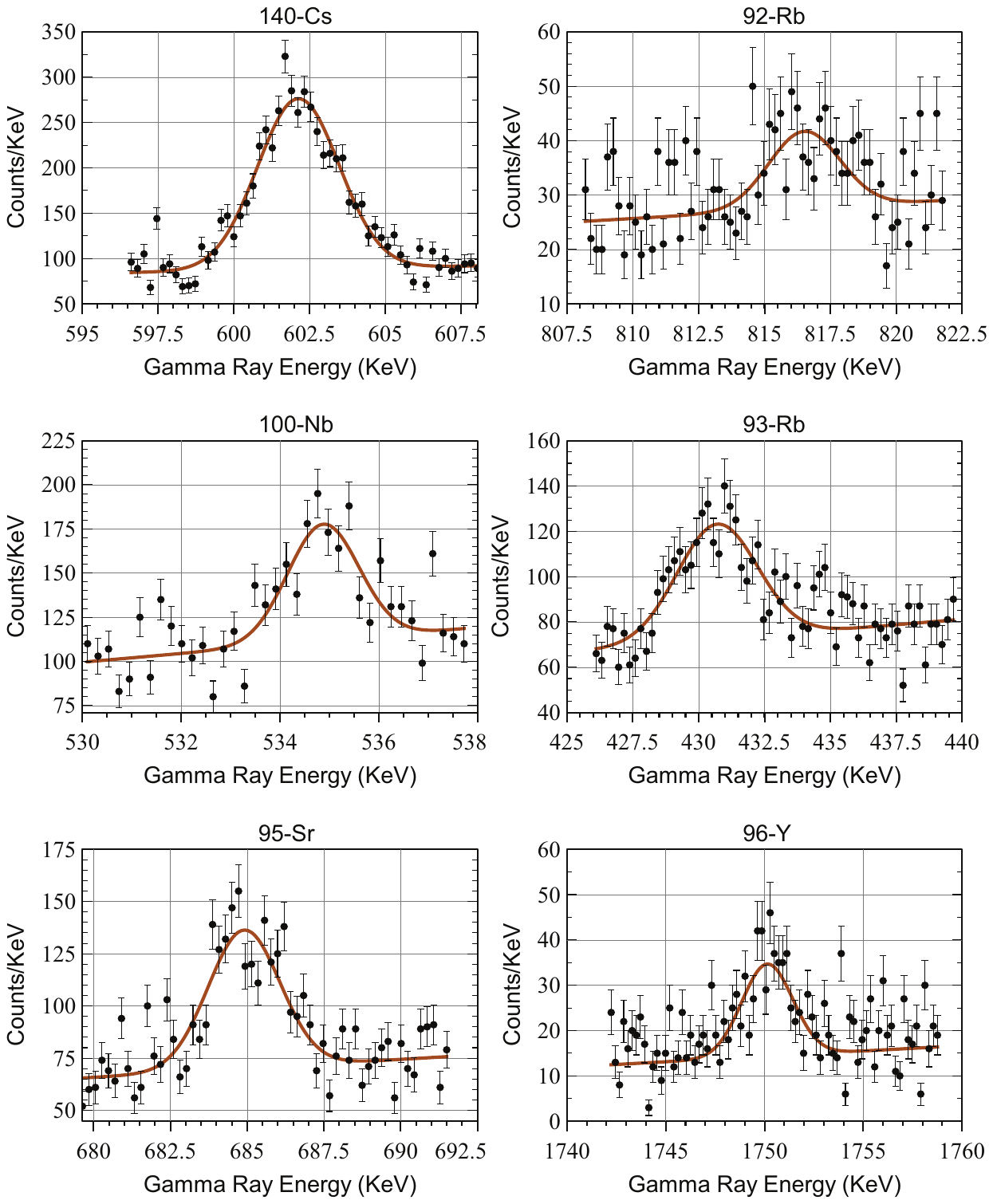}
\caption{\label{fig:fig4} Data and fits for the six gamma-ray peaks of interest from 239-Pu fission  are shown.  The measured spectra are shown by the circled  dots with 1-$\sigma$ uncertainties, and the fits by solid lines. Due to low yield, environmental backgrounds, and Compton scattering from higher energy photons, gamma ray peaks from $\mathrm{^{97}Y}$  and $\mathrm{^{142}Cs}$ are not detectable, and omitted in this figure.  }
\end{figure}

Peaks were analyzed from the measured energy spectra using two methods: non-linear fitting and a simple summation. The ANSI standard for the summation method is given in \cite{768889,MARLAP, EPA402}, and the detailed explanation is given in \cite{knoll2000radiation, gilmore2008practical}. The fit function was a combination of a Gaussian and a linear continuum. Fit analysis was performed using GF3M program from the RADWARE package \cite{radfordradware} and an open-source software, GNUPLOT \cite{williams20041}.   Details for the fitting method are described in the references. Fig. \ref{fig:fig3} and Fig. \ref{fig:fig4} show the data and fits for fission product gamma-ray peaks of interest resulting from neutron irradiation of \(\U \) and \(\Pu \), respectively. 
Table \ref{tab:table5} and Table \ref{tab:table6} summarize the fitting statistics of \(\U \) and \(\Pu \).

In general, the fitted peak energies are consistent with tabulated values for both \(\U \) and \(\Pu \). However, the p-values suggest that a single  Gaussian may be a poor model for some peaks, likely indicating interference from additional unidentified gamma rays.  This could be clarified with greatly improved statistics. As shown in Fig. \ref{fig:fig1}, \(\Pu \) generates generally more gamma-ray activities than \(\U \), suggesting more interference. This fact appears to be consistent with all p-values being lower for \(\Pu \) compared to  \(\U \). 

\begin{table*}[t]
\caption{\label{tab:table5}%
Best-fit values and fitting statistics for the fission products from \(\U \) are summarized. Fitted energies are consistent with the calibrated energies.  But fitted FWHMs show some divergence.  The fit statistics (p-values) indicate acceptable fits for $\mathrm{^{100}Nb}$, $\mathrm{^{140}Cs}$ and $\mathrm{^{95}Sr}$. (see Results section for detail). }
\begin{ruledtabular}
\begin{tabular}{ccccccc}
\textrm{\(\U\)}&
\textrm{$\chi^2$/DOF}&
\textrm{Calibrated Centroid}&
\textrm{Fitted Centroid}&
\textrm{Calibrated FWHM}&
\textrm{Fitted FWHM}&
\textrm{p-value}\\
\colrule
$\mathrm{^{140}Cs}$ &1.63 &602.33 &602.28(4) & 1.93 & 2.87(8) & 0.05\\
$\mathrm{^{95}Sr}$&1.43  &685.57 &685.43(6) & 2.01 & 2.50(11) & 0.03\\
$\mathrm{^{100}Nb}$& 1.26  &535.61 &535.4(3) & 1.87 & 2.4(10) & 0.02\\
$\mathrm{^{93}Rb}$&1.21  &432.67 & 430.38(15) & 1.76 & 3.4(3) &  $<$ 0.01\\
$\mathrm{^{92}Rb}$&2.26  &814.98 &814.6(5) & 2.12 & 1.1(12) &  $<$ 0.01\\
$\mathrm{^{96}Y}$ &1.62  &1750.50 &1749.4(3) &  2.73 & 1.9(6) & $<$ 0.01 \\
\end{tabular}
\end{ruledtabular}
\end{table*}

\begin{table*}[t]
\caption{\label{tab:table6}%
Best-fit values and fitting statistics for the fission products from \(\Pu \) are summarized. Similar to the case of \(\U \), fitted energies are consistent with the calibrated energies while fitted FWHMs show some divergence.  Unlike \(\U \), the fit statistics (p-values) indicate poor fit quality for $\mathrm{^{100}Nb}$, $\mathrm{^{140}Cs}$ and $\mathrm{^{95}Sr}$ due to interference from other gamma rays.  (see Results section for detail).
}
\begin{ruledtabular}
\begin{tabular}{ccccccc}
\textrm{\(\Pu\)}&
\textrm{$\chi^2$/DOF}&
\textrm{Calibrated Centroid}&
\textrm{Fitted Centroid}&
\textrm{Calibrated FWHM}&
\textrm{Fitted FWHM}&
\textrm{p-value}\\
\colrule
$\mathrm{^{140}Cs}$ &1.90 &602.33 &602.03(5) & 1.93 & 3.41(11) & $<$ 0.01\\
$\mathrm{^{95}Sr}$&1.37  &685.57 &685.45(6) & 2.01 & 2.41(11) & $<$ 0.01\\
$\mathrm{^{100}Nb}$& 2.74  &535.61 &534.75(10) & 1.87 & 2.09(23) & $<$ 0.01\\
$\mathrm{^{93}Rb}$&2.92  &432.68 & 434(10) & 1.76 & 5(18) &  $<$ 0.01\\
$\mathrm{^{92}Rb}$&1.89  &814.96 &815.4(3) & 2.12 & 1.9(10) &  $<$ 0.01\\
$\mathrm{^{96}Y}$ &1.9  &1750.50 &1750.19(13) &  2.73 & 2.23(23) & $<$ 0.01 \\
\end{tabular}
\end{ruledtabular}
\end{table*}

\section{Results}

For each gamma-ray, the statistical significance is determined using the method of Refs. \cite{knoll2000radiation, gilmore2008practical, MARLAP}. This method involves two statistical limits: Lc and Ld.  The critical limit (Lc) is defined as the net count of a gamma ray peak above which a sample net count is statistically significant with the probability of false positive given by $\alpha$.  The detection limit (Ld) is defined as the  net count of a gamma ray peak above Lc that has a probability of false negative given by $\beta$. 
We adopt the usual convention of using $\alpha$ = $\beta$ = 0.05 as the desired level of statistical significance. We note that the statistics in this method are based on a one-sided 95\% confidence level so that the z statistic cutoff is 1.65, not 1.96. 
For the non-linear fitting method, Lc and Ld are given by Eq. (\ref{eq4}) and (\ref{eq5})  where $\sigma = \sqrt{B}$ and B = background count (no sample is present) respectively \cite{knoll2000radiation}.

\begin{equation}
 L_\textrm{{c}} = 
 2.33\, \sigma
\label{eq4}
\end{equation}
\begin{equation}
 L_\textrm{{d}} = 
 2.71+4.65\, \sigma
\label{eq5}
\end{equation}

Due to low yield, environmental backgrounds, and Compton scattering from higher energy photons, gamma ray peaks from $\mathrm{^{97}Y}$  and $\mathrm{^{142}Cs}$ are not detectable in the present data. A follow-on  experiment is under consideration, using  larger sample sizes and a shorter RABBIT transit time.
For $\mathrm{^{93}Rb}$, $\mathrm{^{92}Rb}$ and $\mathrm{^{96}Y}$, the fitted net counts  are much larger than the expected. This is suspected to be mainly due to interference from other gamma rays. When fitted net counts are adjusted appropriately to account for interference, they are shown to be consistent with the expected net count.  However, the results are below the statistical significance and are inconclusive. 

The contribution from interference is estimated as described below.  For  $\mathrm{^{93}Rb}$, 6 nuclides (A = 90, 134, 138, 143, 144, 145) produce a similar or larger order of magnitude of gamma-ray yield ($I_{\gamma}$ $\times$ total fission yield) in the 432 keV region in our data \cite{robinson2015radicalc, osti_5352089}. Based on the estimate of the gamma rays having a measurable effect, the proportion of gamma-ray yield of $\mathrm{^{93}Rb}$ with respect to the 6 nuclides gives about 6\% which is consistent with the expected net count of $\mathrm{^{93}Rb}$. In addition, $\mathrm{^{93}Rb}$ (432.61 KeV, $I_{\gamma}$ = 0.202) and $\mathrm{^{143}Ba}$ (431.2 KeV, $I_{\gamma}$ = 0.0276) are expected to produce approximately the same number of counts in our data. The 431.2 keV gamma-ray peak was fitted to obtain its net count which shows consistency with $\mathrm{^{93}Rb}$.
As for $\mathrm{^{92}Rb}$, 13 nuclides (A = 82, 91, 92, 101, 132, 133, 132, 136, 137, 139, 140, 144, 147) produce a similar or larger order of magnitude of gamma-ray yields than $\mathrm{^{92}Rb}$ in the 815 keV energy region \cite{robinson2015radicalc, osti_5352089}. The proportion of  measured peak counts from $\mathrm{^{92}Rb}$ with respect to the thirteen nuclides is about 1.4\% which is consistent with the expected net count of $\mathrm{^{92}Rb}$. 
For $\mathrm{^{96}Y}$ , the strongest gamma ray energy is at 1750.4 KeV with $I_{\gamma}$ = 0.0235, with interference from $\mathrm{^{96m}Y}$  (1750.06 KeV, $I_{\gamma}$ = 0.88).  A proportion of gamma rays yield from $\mathrm{^{96}Y}$ with respect to the total gamma rays yields from both $\mathrm{^{96}Y}$ and $\mathrm{^{96m}Y}$ is about 3\% which is consistent with the expected net count of $\mathrm{^{96}Y}$. 

\begin{figure}[t]
\includegraphics[scale=0.45]{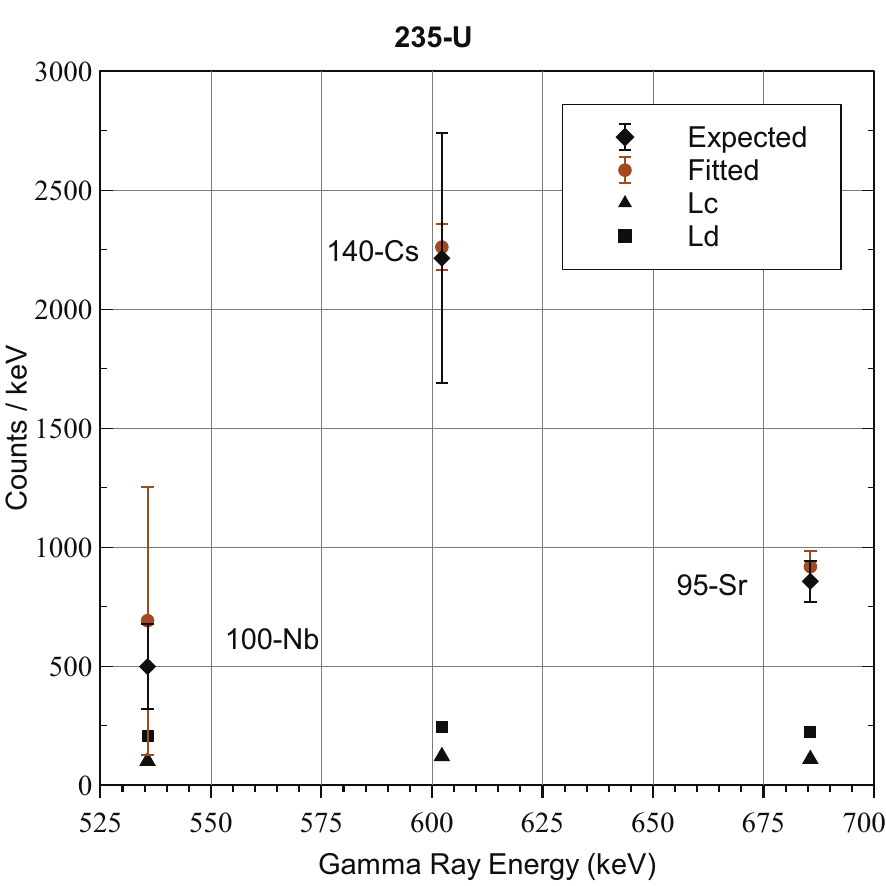}
\caption{\label{fig:fig5} Fitted and expected net counts and statistical limits and uncertainties for \(\U \). The yields of $\mathrm{^{100}Nb}$, $\mathrm{^{140}Cs}$ and $\mathrm{^{95}Sr}$ are shown to be consistent with the expected values.  $\mathrm{^{93}Rb}$, $\mathrm{^{92}Rb}$ and $\mathrm{^{96}Y}$ are below the statistical limit of detection, and are excluded from the plot for clarity. 
}
\end{figure}

\begin{figure}[h]
\includegraphics[scale=0.45]{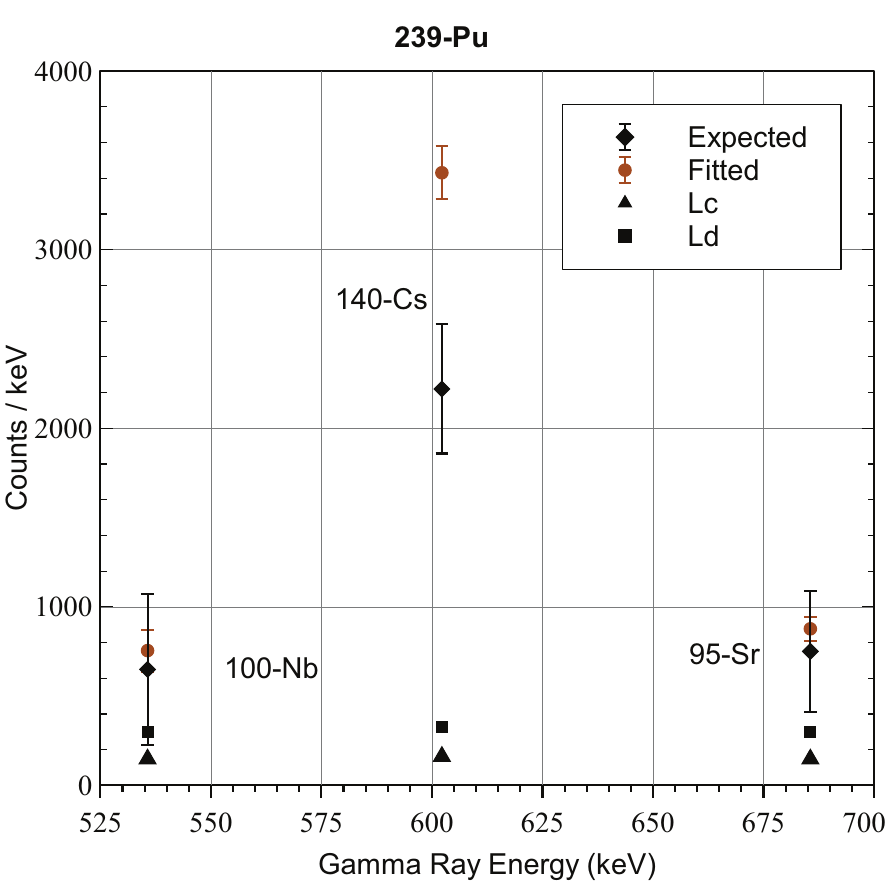}
\caption{\label{fig:fig6}Fitted and expected net counts and statistical limits and uncertainties for \(\Pu \).  The yields of $\mathrm{^{100}Nb}$ and $\mathrm{^{95}Sr}$ are plotted and shown to be consistent with the expected values.  The fitted $\mathrm{^{140}Cs}$ is about 35\% larger than the expected value, suggesting a possible problem with the JEFF3.3 fission yield library. 
$\mathrm{^{93}Rb}$, $\mathrm{^{92}Rb}$ and $\mathrm{^{96}Y}$ are below the statistical limit of detection, and are excluded from the plot for clarity. 
}
\end{figure}

Fig. \ref{fig:fig5} and Fig. \ref{fig:fig6} show the main results of the present study.  The net peak counts, the limits (Lc and Ld), and the expected counts calculated from the JEFF3.3 fission yields and the detector simulations, are plotted vs. gamma ray energy.

For both $\mathrm{^{235}U}$ and $\mathrm{^{239}Pu}$, the measured gamma rays for $\mathrm{^{100}Nb}$ and $\mathrm{^{95}Sr}$ are statistically significant ($\alpha$ = 0.05), are above the minimum detection limit ($\beta$ = 0.05), and are fully consistent with the expected counts. This indicates that the independent fission yield from JEFF3.3 fission yield library and gamma-ray intensity from ENDF/B-VIII.0 decay data sublibrary are reliable for these nuclides.
For $\mathrm{^{140}Cs}$, the measured gamma ray yield is  statistically significant, and for the case of \(\U \) its value is consistent with that expected. But  for the case of \(\Pu \), the measured count is 35\% larger than the expected value.  This suggests a possible problem with the fission yield value in JEFF3.3, which should be confirmed by a follow up study.  
A possible complication for the present type of measurement was pointed out by Hayes et al. \cite{hayes2016reactor} in the form of a potential contribution from epithermal neutron induced fission.  The measured epithermal neutron flux at HFIR for this study is  0.4\% (0.7\%) of the thermal neutron flux for  \(\U \) ( \(\Pu \)). In this analysis, the contribution  from epithermal neutrons is too low to make any significant difference in the data.  The role of epithermal neutrons should be further investigated in an actual reactor environment where the fuel composition is precisely known.

\section{Conclusion}

The summation method \cite{Muller-PRC054615} used to estimate the $\bar\nu_e$ spectrum analysis depends on accurate data values in the nuclear libraries. To check the fission yield library values, $\mathrm{^{235}U}$ and $\mathrm{^{239}Pu}$  samples are irradiated using HFIR, and gamma-ray spectroscopy is used evaluate the gamma-rays from 8 short-lived fission products which were suggested\cite{dwyer2015spectral} as a possible source of the spectral bump in the reactor neutrino spectrum at the 5 to 7 MeV region .   

The gamma ray yields for $\mathrm{^{100}Nb}$ and $\mathrm{^{95}Sr}$ from both $\mathrm{^{235}U}$ and $\mathrm{^{239}Pu}$, as well as $\mathrm{^{140}Cs}$  from $\mathrm{^{235}U}$ are found to be statistically significant and consistent with expectation based on the JEFF3.3 fission yield library. However, an inconsistent result was found for $\mathrm{^{140}Cs}$  from $\mathrm{^{239}Pu}$, which suggests that the JEFF3.3 fission yield value for this nuclide may be incorrect. 

The results for remaining fission products are inconclusive due to insufficient statistics.  A follow-on experiment with increased sample sizes and a faster reactor-to-counting station transfer is in discussion.
Overall, the present study underscores the importance of continuous improvement and refinement of nuclear data libraries. 
Additional experimental data and continued analysis of existing data are important for verifying and improving fission yield values.

\begin{acknowledgments}
Authors would like to thank A. Sonzongni and E. McCutchan for valuable comments.
This work was funded by the U. S. National Science Foundation under grant NSF-PHY1242611, NSF-1812504, 1747523, and 1314483. The work at Brookhaven National Laboratory was sponsored by the Office of Nuclear Physics, Office of Science of the U.S. Department of Energy under Contract No. DE-AC02-98CH10886 with Brookhaven Science Associates, LLC.

\end{acknowledgments}

\newpage
\nocite{*}

\bibliography{apssamp}

\end{document}